\documentclass[aps,prc,twocolumn]{revtex4}
\usepackage{graphicx}
\usepackage{amsmath,bbm}
\usepackage{amssymb,bm}
\usepackage{array}
\newcommand\nd{\noindent}
\begin{document}
\title{ A proposal for exploring quantum theory in curved spacetime in lab}
\author{S.~Ganesh\footnote{Corresponding author:\\Email: gans.phy@gmail.com}}
\affiliation{Sri Sathya Sai Institute of Higher Learning, Prasanthi Nilayam - 515134, A.P., India}
\begin{abstract}
	Gravity curves spacetime. In regions where the de Broglie wavelength is very small compared to the curvature of spacetime, the wave equations in flat spacetime can be generalized to curved spacetime. The validity of the formulation  when the de Broglie wavelength becomes comparable to the curvature is an open question.
To test these formulations experimentally, huge energy of the order of the Planck mass would be required. 
Euclideanized spacetime is used to model thermal systems.
	In this work, an equivalence between spatial variation of temperature in thermal baths and curvature of Euclideanized spacetime is propounded. The variation in temperature is recast as a variation in the metric. The Dirac equation is then solved in this curved Euclideanized spacetime. The curvature in Euclideanized spacetime can be obtained in Chromodynamic or Electromagnetic scale energies.
The equivalence has limitations, but nevertheless, if correct, it could provide a platform to experimentally explore quantum mechanics in curved spacetime.

\vskip 0.5cm

{\nd \it Keywords } : Quantum theory, Curved spacetime, Dirac Equation, Temperature, Equivalence \\
{\nd \it PACS numbers } : 04.62.+v, 04.60.-m, 04.20.Gz  

\end{abstract}

\maketitle
\section{Introduction}
\label{sec:intro}
  Quantum theory in curved spacetime is a topic of intense research.
%  ~\cite{book1, mukh, padh1, padh2, arb1}. 
  In fact, it has led to many interesting phenomenon like the Hawking radiation~\cite{hawking} and Unruh effect~\cite{unruh}. However, testing the quantum theory in curved spacetime is a formidable challenge due to the high energies required to curve spacetime. 
  The regime where the de Broglie wavelength becomes comparable to the spacetime curvature is an even more interesting regime, as the current physics is expected to break down.
  However, experimental exploration in this regime is an even more formidable challenge, as the energies involved are in the Planck scale.

  On a different front, Euclideanized spacetime is often used to model thermal systems. The temperature is modeled as inverse of imaginary time. In this work, the equivalence between temperature and Euclideanized spacetime is taken a step further, by modeling the temperature variation in space as curvature in Euclidean spacetime.
  It may be clarified that a spatial variation in temperature does not curve the original Minkowski spacetime. It is the mathematical equivalence between the spatial variation of temperature and the curvature of the corresponding Euclideanized spacetime that is being propounded here.
  If this equivalence is indeed true, then experimental results in a system of a thermal bath, with temperature gradient, can be used to explore the formulation of quantum mechanics in curved spacetime.
  
  If the Euclideanized space-time is curved, then it should be possible to formulate a geodesic equation. The geodesic equation and its interpretation in terms of kinetic theory is developed in Sec.~\ref{sec:geodesic}.
  The geodesic curves have limitations compared to their curved Minkowski counterparts. Only, the spatial variation of temperature has been captured while formulating the metric. Thus the geodesics can only curve in spatial dimensions.
  The geodesic equation gives an insight into the nature of the equivalence.
  We also elucidate the effect of the curvature on a fermion or  boson wavefunction, when it passes through this Euclideanized curved space-time in Sec.~\ref{sec:dirac} and~\ref{sec:kleingordon}. The effects are not identical to a curved Minkowski space-time, nor is it expected to be. In the imaginary time formalism, the time variable is replaced by inverse temperature, and thus the explicit time variable is absent in this formulation. In the absence of the time variable, the frequency of a photon would remain unaffected. However, the Matsubara frequencies do get affected, in an equivalent way, by the Euclideanized space-time curvature. 
  The effects of the Euclideanized space-time curvature would then be the effects of modification of the Matsubara frequency. The key would be the identification of the mathematical equivalence between the two curved space-times.

  The  thermal bath can either be a Quantum Chromodynamic (QCD) plasma or a Quantum Electrodynamic (QED) plasma.
 If the thermal bath were to be a QCD plasma, namely the Quark Gluon Plasma (QGP), then the energies involved would be in the GeV scale and the distances involved would be in femtometer (fm) scale. These are realistic systems that are being produced at the Large Hadron Collider (LHC) and Relativistic Heavy Ion Collider (RHIC)~\cite{nature}. 
 If however the thermal bath were to be a QED plasma, then the energies reduce further to keV or even tens of ev~\cite{plasma1}, and the distances involved would be in nanometer (nm) or picometer (pm) range.
 A QED plasma is composed of electrons, ions and neutral atoms. There are typically two types of QED plasma, a) thermal plasma, which is in local thermal equilibrium and b) non-thermal or cold plasma~\cite{plasma2}. 
 In a thermal plasma, all the species constituting the plasma are at the same or similar temperature~\cite{plasma2}.
 %gans: april 2020
 A thermal plasma would be expected to be more appropriate for exploring quantum effects, as the temperature is more well defined. 
% Since the quarks also carry an electric charge, the QGP might also be treated as a QED plasma. However, because of much higher energies involved and shorter lifetimes (about 5 fm/c or so~\cite{gans2,gans3}), it may have negligible effect on QED particles like electron etc.

  The rest of the paper is as follows.  In Sec. \ref{sec:polyakov}, we first look at the Wilson Polyakov loop correlator which determines the potential between two particles in a thermal bath. This is used to motivate the equivalence between temperature gradients and curvature of spacetime. 
  The geodesic equations, in a classical context, are explored in ~\ref{sec:geodesic}.
  In Sec. \ref{sec:dirac}, we solve the Dirac equation in the Euclidean curved spacetime for a massive spinor. This section also captures the numerical work performed for solving the Dirac equation in presented. We also explore how the fermion 3-momentum gets affected by the space-time curvature. In Sec~\ref{sec:kleingordon}, we touch upon the  Klein Gordon particle, namely the photon.  Finally, in Sec. \ref{sec:conclusion}, we look at the conclusions.

%gans
\section{The Equivalence}
\label{sec:polyakov}
\begin{figure}
\includegraphics[width = 80mm,height = 80mm]{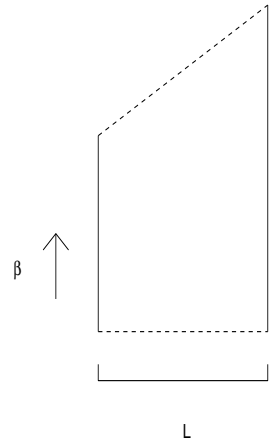}
\caption{Wilson-Polyakov loop to model the potential between two particles in a potential gradient.
	The two solid vertical lines represent the two Wilson lines corresponding to the two Polyakov loops $P(0)$ and $P(L)$.}
\label{fig:single_loop}
\end{figure}

Let us first consider a scenario where  a heavy Quark and anti-Quark are situated in a thermal bath, having a temperature gradient.
The Wilson-Polyakov loop that has been used to model the scenario of the temperature gradient is depicted in Fig.~\ref{fig:single_loop}. 
The relation of the Wilson-Polyakov loop correlator to the potential between the Quark and anti-Quark could be modeled as~\cite{soo}:   
\begin{equation}
\label{eq:qqbeta}
%gans
%<W(C)> = e^{-V_{Q\bar{Q}}\beta},
<P(0)P(L)> = e^{-V_{Q\bar{Q}}\beta},
\end{equation}
where, $\beta = \frac{1}{\Theta}$ 
and $\Theta$ is the temperature of the system. $V_{Q\bar{Q}}$ is the potential between the Quark and anti-Quark. Here, in case of a QCD system, the two particles in the system are the Quark and anti-Quark. One can however chose a QED system, in which, the particles in the thermal bath could be  two oppositely charged leptons, e.g., muon and anti-muon. The Polyakov loop is the trace of the Wilson line integrated from 0 to $\beta$ in Euclidean time, i.e., $P(x) = \frac{1}{N}Tr\Big [T exp(-i\int_0^{\beta}A_0(x,\tau)d\tau \Big ]$. Here, $A(x,\tau)$ is the pure gauge field. In the case of QCD, a time ordering operator, $T$, would be required. 
While evaluating the Polyakov loop correlator, the Minkowski space time is converted to an Euclidean time by the use of imaginary periodic time.

If there is a temperature gradient, then the R.H.S. of Eq.~\ref{eq:qqbeta}, poses a problem. There is no provision to have a spatial variation of $\beta$. To resolve this issue, let us probe the Euclideanized spacetime further.
%epjc
In the imaginary time formalism, the spacetime fabric is a uniform topological cylinder in a Euclideanized space of radius $\frac{\beta}{2\pi}$. The contour in Fig. \ref{fig:single_loop} seems to indicate that the cylinder is non-uniform, with radii $\frac{\beta}{2\pi}$ and $\frac{\beta + \Delta \beta}{2\pi}$ at the two ends. Thus application of Eq.~\ref{eq:qqbeta} becomes a challenge. However, it is possible to alternatively view the topological cylinder as a uniform cylinder in a curved Euclidean space (a straight line in curved space would be curved). 
 If one were to evaluate the Polyakov loop correlator, in Eq.~\ref{eq:qqbeta}, in the field theoretic domain, one would then in principle, evaluate the correlator in the curved Euclidean space, with a "constant" $\beta$. 

 Another perspective is that in a Minkowski spacetime, if time were to depend on the spatial location, then, the dependence can be seen as a result of curvature in the Minkowski spacetime. Analogously, if temperature were to depend on spatial location, then the Euclideanized spacetime should be curved. 

 This gives rise to an equivalence principle, that variation in temperature can be recast as a variation in the metric.
 If this is indeed true, it is possible to mimic curved Euclideanized spacetime by resorting to spatial variation of temperature in a thermal bath. 

 Since spatial temperature variation is required, the thermal bath (either the QCD or QED thermal bath), would be in a state of local thermal equilibrium instead of a full-fledged thermal equilibrium.
 It seems that QGP system can be described by ideal hydrodynamics~\cite{naturephy},  which implies that the QCD plasma would be in local thermal equilibrium. 
 As described in Sec. \ref{sec:intro}, the QED plasma can be either a cold plasma or a thermal plasma. The thermal plasma is hotter, and is in a state of local thermal equilibrium~\cite{plasma2}. The cold plasma is however, a non-equilibrium thermal state.
 The formalism presented here, would then apply mainly to the QGP or the QED thermal plasma, which is in a state of local thermal equilibrium.
%epjc

 Thus we see that by drawing upon the equivalence between temperature gradients and Euclideanized spacetime curvature, it is possible to study curved spacetime at energies as low as keV scale. Temperature gradients in GeV scale are already produced in the lab at the LHC and RHIC. However, these are QCD systems in the strong coupling regime, and hence difficult to model analytically. QED system in the keV range may present an excellent opportunity to study experimentally, curved Euclideanized spacetime, and as well as model it analytically to a high degree of accuracy.

\section{Geodesic Equation}
\label{sec:geodesic}
Let us consider a candidate metric, where the temperature varies across the two dimensional $x-z$ plane.
\begin{equation}
\label{eq:metric2D}
	ds^2 = g_{00}(x,z)dt^2 - dx^2 - dy^2 - dz^2,
\end{equation}
with $g_{00}(x,z) = V(x,z)$, and $t = i\beta$.
The temperature variation is given by $\Theta(x,z) = \frac{1}{\sqrt{V(x,z)}}$.

The set of geodesic equations for the above metric are
\begin{eqnarray}
	\label{eq:geodesic}
	%equation 1
\nonumber	\frac{\partial\beta}{\partial\lambda} \frac{\partial V(x,z)}{\partial z}\frac{\partial  z}{\partial \lambda} + 
	\frac{\partial\beta}{\partial\lambda} \frac{\partial V(x,z)}{\partial x}\frac{\partial  x}{\partial \lambda} + 
	\frac{\partial^2 \beta}{\partial \lambda^2} V(x,z) = 0, \\
	%equation 2
\nonumber	\left (\frac{\partial \beta}{\partial \lambda} \right )^2 \frac{\partial V(x,z)}{\partial x}  - 2\frac{\partial^2 x}{\partial \lambda^2} = 0,\\
	%equation 3
\nonumber	\frac{\partial^2 y}{\partial \lambda^2} = 0,\\
	%equation 4
	2\frac{\partial^2 z}{\partial \lambda^2} - \left ( \frac{\partial \beta}{\partial \lambda} \right )^2 \frac{\partial V(x,z)}{\partial z} = 0,
\end{eqnarray}
where, $\lambda$ is an affine parameter.
Eliminating $\left (\frac{\partial \beta}{\partial \lambda} \right )^2$ between the second and fourth of Eqs \ref{eq:geodesic}, 
\begin{equation}
	\label{eq:xzcurve}
	\frac{\partial V}{\partial x} \frac{\partial^2 z}{\partial \lambda^2} = \frac{\partial V}{\partial z} \frac{\partial^2 x}{\partial \lambda^2}.
\end{equation}
Equation~\ref{eq:xzcurve} indicates that for any non-zero $\frac{\partial V}{\partial x}$ and $\frac{\partial V}{\partial z}$, the particle will bend, and the geodesic is not straight.
For photons, the above can be seen equivalent to bending of light in a medium of varying refractive index. 
Since, no quantum condition has been imposed in arriving at the geodesic equation, it is expected to be valid in the classical limit, where the deBroglie wavelength of the particle is much smaller than the space-time curvature.
\subsection{Interpretation at a classical level}
We now look at the possible origins of the geodesic curve, given by Eq. \ref{eq:xzcurve}, from a kinetic theory point of view for a massive particle.
We study the propagation of a massive test particle in a plasma.
Let $\langle v^2(x,z)\rangle$ be the expected r.m.s velocity of a constituent particle of the medium at spatial location (x,z).
Classically,
\begin{equation}
	\frac{1}{2}m_p\langle v^2 \rangle = \frac{n_p}{2}k\Theta,
\end{equation}
where $n_p$ and $m_p$ are the degrees of freedom and the effective mass of the constituent plasma particle, and $\Theta$ is the temperature.
This means,
\begin{equation}
	\sqrt{\langle v^2(x,z) \rangle } = \sqrt{\frac{n_p k}{m_p}} \sqrt{\Theta} = k'\sqrt{\Theta}.
\end{equation}
Rewriting in terms of the metric $g_{oo} = V(x,z)$,
\begin{equation}
	\sqrt{\langle v^2(x,z) \rangle } = \frac{k'}{V(x,z)^{1/4}}.
\end{equation}
The gradient in the velocity along the $x$-direction, would then become,
\begin{equation}
	\partial_x\sqrt{\langle v^2(x,z) \rangle } = \frac{k''}{V(x,z)^{5/4}}V_x,
\end{equation}
where, $V_x = \frac{\partial V(x,z)}{\partial x}$.
This variation in r.m.s velocity of the plasma constituents will bring about a change in momentum of the test particle in the plasma.  
The momentum change of the test particle over a unit distance $\Delta x = \Delta l$, is given by 
\begin{equation}
	\Delta p_x \propto \partial_x\sqrt{\langle v^2(x,z) \rangle } \Delta l
\end{equation}
Acceleration of the test particle, would be proportional to the rate of change of momentum. This gives, $a_x  \propto  \frac{\Delta p_x}{\Delta t}$. Under conditions of local thermodynamic equilibrium and stationarity, $\partial_x\sqrt{\langle v^2(x,z) \rangle }$ is static in time.
Furthermore, the time variable $t$, is replaced by the affine parameter $\lambda$ (i.e. $\frac{\partial}{\partial t} \propto \frac{\partial}{\partial \lambda}$).  This finally gives,
\begin{equation}
	\label{eq:accx}
	a_x  \propto \partial_x\sqrt{\langle v^2(x,z) \rangle } \frac{\Delta l}{\Delta \lambda} = \frac{k'' V_x}{V(x,z)^{5/4}} \frac{\Delta l}{\Delta \lambda}
\end{equation}
In the $z$-direction, we again calculate the momentum change $p_z$ over the same (in magnitude) unit distance $\Delta z = \Delta l$. This gives,
\begin{equation}
	\label{eq:accz}
	a_z  \propto  \frac{k'' V_z}{V(x,z)^{5/4}} \frac{\Delta l}{\Delta \lambda}
\end{equation}
Rewriting, $a_x = \frac{\partial^2x}{\partial \lambda^2}$ and $a_z = \frac{\partial^2z}{\partial \lambda^2}$, and taking the ratio of Eqs.~\ref{eq:accx} and~\ref{eq:accz}, we get,
\begin{equation}
	\frac{1}{V_x}\frac{\partial^2 x}{\partial \lambda^2} = \frac{1}{V_z}\frac{\partial^2 z}{\partial \lambda^2},
\end{equation}
which is nothing but the geodesic equation, Eq.~\ref{eq:xzcurve}.

   What we have achieved here is that the curved path taken by the test particle determined by kinetic considerations has been reinterpreted as a geodesic in the Euclideanized space-time.

\section{Dirac Spinor}
\label{sec:dirac}
\subsection{Formulation}
\label{sec:diracform}
A candidate Euclideanized metric representing spatial temperature variation along the 1 dimensional z direction is:
\begin{equation}
\label{eq:metric}
	ds^2 = g_{00}(z)dt^2 - dx^2 - dy^2 - dz^2,
\end{equation}
with $g_{00}(z) = V(z)$, and $t = i\beta$.

The Ricci scalar curvature for the metric in Eq.~\ref{eq:metric} is given by
\begin{equation}
\label{eq:ricci}
	R = \frac{2VV'' - V'^2}{2V^2},
\end{equation}
where  the $'$ refers to differentiation w.r.t $z$.

The Dirac equation in curved space time is given by~\cite{dirac, dirac_rus, book2}:
\begin{equation}
\label{eq:dirac}
	i\gamma^a e^{\mu}_a D_{\mu}\psi - m\psi = 0,
\end{equation}
where,
\begin{itemize}
\item \begin{math}
	D_{\mu} = \partial_{\mu} - \frac{i}{4}\omega_{\mu}^{ab}\sigma_{ab},
\end{math}
\item $\omega_{\mu}^{ab}$ is the spin connection,
\item $\sigma_{ab} = \frac{i}{2}[\gamma_a,\gamma_b]$, and $\gamma_a$ being the Dirac matrices in flat spacetime, and 
\item $e_{\mu}^a$ is the vierbein.
\item In vacuum (0 temperature), $m$ is the gravitational mass. But in the proposed thermal system, an effective mass $m$ has been used. 
\end{itemize}
	In the notation used, $\mu$, $\nu$, etc., are indices in curved spacetime, and $a$, $b$, etc., are indices in flat spacetime.

	The mass $m$, is not the gravitational mass $m_g$, but an effective mass, $m = m_g - i\gamma^0 E$, where $E$ is the energy  eigenvalue of a stationary state. The effective mass has been so chosen, so that this equation boils down to the Dirac Equation in vacuum for a stationary plane wave $exp\{-i(Et - \vec{p}.\vec{x})\}$, when the inverse  temperature $\beta \rightarrow \infty$.
    While describing a Klein Gordon particle in Sec.\ref{sec:kleingordon}, we show that the use of effective mass reproduces the dispersion relation for the photon.

%%%   \subsection{Massless Fermion}
%%%   		At very high energies, we can treat the light fermions as massless. For a massless fermion, the Dirac equation \ref{eq:dirac} becomes
%%%   \begin{equation}
%%%   	\label{eq:massless_dirac}
%%%   	i\gamma^a_e^{\mu}_a\partial_{\mu}\psi = 0,
%%%   \end{equation}
%%%   where we have replaced the covatiant derivative by the partial derivative.
%%%   We take the spinor wavefunction $\psi$ to be of the form
%%%   \begin{equation}
%%%   	\label{eq:massless_psi}
%%%   	\psi = u(z) exp{i(g_{00}\p^0\beta)}
%%%   \end{equation}
%%%   
%%%   		The equation \ref{eq:massless_psi} can be easily be seen to solve Eq.~\ref{eq: massless_dirac} for $\frac{\partial  q(z)}{\partial z} = \frac{p^0}{\sqrt{V(z)}}$ and 
%%%   		\begin{math}
%%%   			u = \begin{table}
%%%   				\begin{tabular}{c}
%%%   					\zeta1\\
%%%   					\zeta2\\
%%%   				\end{tabular}
%%%   				\end{table}, 
%%%   		\end{math}
%%%   		with, $\zeta1 = 
%%%   		\begin{table}
%%%   			1\\
%%%   			0\\
%%%   		\end{table}$ or 
%%%   		$\begin{table}
%%%   			0\\
%%%   			1\\
%%%   		\end{table}$. 
%%%   		Thus, the z-momentum  $\frac{\partial q(z)}{\partial z}$ is seen to vary based on the spatial location. This variation can be experimentally verified to see if the massless Dirac Eq \ref{eq:massless_dirac} is the right equation to describe a  massless fermion. 
%%%   
%%%   		The most interesting case however arises  when the de Broglie wavelength $\lambda \approx R$, where R is the Ricci scalar curvature given by  Eq.~\ref{eq:ricci}.
%%%   		To study, this case, we need to  study  a slowly moving massive fermion.
		We now solve the Dirac equation given by Eq.~\ref{eq:dirac}. 
		The coordinate axes used are $(i\beta,x,y,z)$. The term, $\frac{-i}{4}\omega_{\mu}^{ab}\sigma_{ab}$, for the metric in Eq.~\ref{eq:metric} evaluates to $\frac{f(z)}{8}\gamma^0\gamma^3$, with $\gamma^{0},~\gamma^3$ being the Dirac matrices in flat space, and $f(z) = \frac{-2V'(z)}{\sqrt{V(z)}}$.
We take the solution to be of the form
\begin{equation}
	\label{eq:massive_psi}
	\psi = u(z) exp(-ig_{00}p^0\beta), 
\end{equation}
		where the spinor $u(z)$ is given by
		\begin{equation}
			\label{eq:spinor_u}
			u(z) = \left( \begin{array}{c} (\zeta_1(z) + i\eta_1(z)) \zeta_0 \\ (\zeta_2(z) + i\eta_2(z)) \zeta_0 \end{array} \right), 
		\end{equation}
		with, $\zeta_0$ = 
		$\left( \begin{array}{c}
			1\\
			0\\
		\end{array} \right )$ or 
		$\left (\begin{array}{c}
			0\\
			1\\
		\end{array} \right )$, i.e. $\zeta_0$ indicates a spin up or spin down particle. 
Substituting in Eq.~\ref{eq:dirac}, and equating both the real and imaginary parts to 0, we get:
\begin{eqnarray}
\label{eq:partialsl}
\nonumber	-k_2 u_1 -  s u_1'  + im_g u_2 = 0,\\
\nonumber	-k_1 u_2 +  s u_2'  + im_g u_1 = 0,\\
%\nonumber	-k_2\zeta_1 -  s\zeta_1'  - m\eta_2 = 0,\\
%\nonumber	k_2\eta_1 +  s\eta_1'  - m\zeta_2 = 0,\\
%\nonumber	-k_1\zeta_2 +  s\zeta_2'  - m\eta_1 = 0,\\
%		k_1\eta_2 -  s\eta_2'  - m\zeta_1 = 0,
\end{eqnarray}

where, 
\begin{itemize}
	\item $u_j = \zeta_j + i\eta_j$; j = 1,2,
	\item $s$ is the eigenvalue of the Pauli $\sigma^3$ matrix. Explicitly, $s = 1$, if $\zeta_0 = \left ( \begin{array}{c} 1\\ 0 \end{array} \right )$ and  s = -1 if $\zeta_0 = \left ( \begin{array}{c} 0\\ 1 \end{array} \right )$. Thus, it's value is twice the spin of the spinor particle. 
	\item $k_1 = iE + \frac{-C - f(z)s/8}{\sqrt{V(z)}}$.
	\item $k_2 = iE + \frac{-C + f(z)s/8}{\sqrt{V(z)}}$.
\end{itemize}
It is to be noted that $m_g$ is the gravitational mass.
In coming up with the equations in Eq.~\ref{eq:partialsl}, we have taken $p^0 = \frac{C}{V(z)}$, where, $\frac{C}{V(0)} = \frac{(2n+1)\pi}{\beta_T(0)}$, n = 0,1,2,...; $\beta_T(z)$ is the inverse temperature.
In flat spacetime, with $V(z) = V_0=$ constant, $\frac{C}{V_0}$ would be the Matsubara frequency. 
%In fact, without loss of generality, one can take $V_0=1$ and $V(0)=1$ (by absorbing the scaling factor  into $\beta$),  which gives, $C = \frac{(2n+1)\pi}{\beta_T(0)}$.
In this article, $C$ itself is sometimes referred to as Matsubara frequency. 
This choice of $p^0$, keeps the exponent in Eq.~\ref{eq:massive_psi} independent of $z$, and is also seen to solve the Dirac equation.
The anti-periodicity of $\psi$ is discussed in Sec.~\ref{sec:matsubara}.

A simple algebraic manipulation of the equations in Eq.~\ref{eq:partialsl}, results in: 
%\nonumber	u_1'' + s(k_2 - k_1)u_1' + (-k_1 k_2 + sk_2' - m_g^2)u_1 = 0,\\
%u_2'' + s(k_2 - k_1)u_2' + (-k_1 k_2 - sk_1' - m_g^2)u_2 = 0.
\begin{eqnarray}
\label{eq:secondorder}
	\nonumber	u_1'' + s\frac{f}{4\sqrt{V}}u_1' + (-k_1 k_2 + sk_2' - m_g^2)u_1 = 0,\\
	u_2'' + s\frac{fs}{4\sqrt{V}}u_2' + (-k_1 k_2 - sk_1' - m_g^2)u_2 = 0.
\end{eqnarray}
  \subsection{Matsubara frequency}
  \label{sec:matsubara}
%	The Matsubara frequency, $p^0 = \frac{C}{V(z)}$, is now analyzed.
	We now explore the anti-periodicity of the fermionic wavefunction $\psi$ around the topological cylinder at all points in space.
	As discussed earlier, let $C = \frac{(2n+1)\pi}{\beta_T(0)}$.
Then, $exp\left \{-ig_{00}(z)p^0(z)\beta  \right \}  = exp\left \{-i(2n+1)\pi \right \} $  at $\beta = \beta_T(0)$, which is independent of $z$. In other words, the anti-periodicity of the fermion wavefunction is retained at all values of $z$. 
	This also determines, that, $\beta = \beta_T(0)$ would be the value used in R.H.S. of Eq.~\ref{eq:qqbeta}. We note that the inverse temperature at a point $z$ is given by "proper" $\beta$ = $\beta_p = \sqrt{g_{00}(z)}\beta$. A "proper" Matsubara frequency could then be given as, $\omega_p = \sqrt{g_{00}}p^0 = \frac{C}{\sqrt{V(z)}}$.

\subsection{Thermal Bath with zero temperature gradient}
To gain an insight into Eq.~\ref{eq:secondorder}, we first solve Eq.~\ref{eq:secondorder} in flat space i.e. when the temperature gradient is 0. 
This would provide a baseline $u_1^{flat}$ and $u_2^{flat}$. 
When, the gradient $V'(z)=0$, and $V(z) = V_0$, the equations in Eq.~\ref{eq:secondorder} simplify to
\begin{eqnarray}
	\label{eq:nograd}
	\nonumber	u_1'' +  ( -(iE -\frac{C}{\sqrt{V_0}})^2  - m_g^2)u_1 = 0,\\
		u_2'' +  ( -(iE -\frac{C}{\sqrt{V_0}})^2  - m_g^2)u_2 = 0,
%\nonumber	\zeta_1'' +  (-\frac{C^2}{V_0}  - m^2)\zeta_1 = 0,\\
%\nonumber	\eta_1'' +  (-\frac{C^2}{V_0}  - m^2)\eta_1 = 0,\\
%\nonumber	\zeta_2'' +  (-\frac{C^2}{V_0}  - m^2)\zeta_2 = 0,\\
%	\eta_2'' +  (-\frac{C^2}{V_0}  - m^2)\eta_2 = 0,
\end{eqnarray}
	whose solution is given by 
	\begin{eqnarray}
		\nonumber		u_1^{flat} = A_1exp\Big ( \frac{1}{\sqrt{2}} \Big \{\sqrt{-Q_r + \sqrt{Q_r^2 + Q_i^2}  }\\
\nonumber		+i\sqrt{Q_r + \sqrt{Q_r^2 + Q_i^2} } \Big \}z  \Big )\\ 
\nonumber		+	 A_2exp\Big ( \frac{-1}{\sqrt{2}} \Big \{ \sqrt{-Q_r + \sqrt{Q_r^2 + Q_i^2}  }\\
		+i\sqrt{Q_r + \sqrt{Q_r^2 + Q_i^2} } \Big \}z \Big ) 
	\end{eqnarray}
where,
\begin{description}
	\item $Q_r =  ( E^2 - m_g^2 ) - \frac{C^2}{V_0}  $ and,
	\item $Q_i = \frac{2EC}{\sqrt{V_0}}$.
\end{description}
	As the temperature  $\Theta \rightarrow 0$, $V_0 \rightarrow \infty$, the resulting solution $ exp^{\left  (i\sqrt{Q_r}z \right )} \rightarrow e^{\left \{i \left (\sqrt{E^2 -  m_g^2}\right )z \right \}}$. This is nothing but the spatial part of the plane wave $e^{-i(Et - \vec{p}.\vec{x})}$, namely, $e^{i\vec{p}.\vec{x}}$, where $\vec{p}$, is the 3-momentum and $p^2= (E^2 - m_g^2)$.
	For finite temperatures, one can infer the following two observations
	\begin{itemize}
		\item The resultant 3-momentum in the thermal bath = $q_i = \sqrt{\frac{1}{2}(Q_r + \sqrt{Q_r^2 + Q_i^2})}$, is less than the 3-momentum in vacuum i.e. $\sqrt{  (E^2 - m_g^2)}$.
		\item There is a decay factor = $q_r = \sqrt{\frac{1}{2}(-Q_r + \sqrt{Q_r^2 + Q_i^2})}$.
	\end{itemize}
The fermion then acts as a particle, with energy $E$, 3-momentum $q_i$ and mass $\sqrt{E^2 - q_i^2}$.

%	Before we go to the section on temperature gradient, it is prudent to make a note on the interpretation of the 1st and second derivative.
%From Eq.~\ref{eq:partialsl}, it can be seen that the $1^{st}$ derivative,
%namely $\frac{u_1^{'}}{u_1}$ or $\frac{u_2^{'}}{u_2}$ mixes the spinor components $u_1$ and $u_2$. 
%Thus, in general, $\frac{u_1^{'}}{u_1}$ or $\frac{u_2^{'}}{u_2}$ need not directly provide the 3-momentum observable, unless the two spinor components $u_1$ and $u_2$ are related by a constant factor. 

It may be possible then to reduce  Eq.~\ref{eq:secondorder} to the form of Eq.~\ref{eq:nograd} i.e. without the first derivative term, $\frac{fs}{4\sqrt{V(z)}}$. 
In the form present in Eq.~\ref{eq:nograd} the second derivative can be used to provide the square magnitude of the 3-momentum. 
This would be the subject matter of discussion in Sec.\ref{sec:tempgrad}.
As a reminder, we have seen that in case of nil temperature gradient, the $\frac{fs}{4\sqrt{V(z)}}$ term disappears, and $q_i = Imag(\sqrt{\frac{u_1^{''}}{u_1}})$ (=  $Imag(\sqrt{\frac{u_2^{''}}{u_2}})$) directly provides the 3-momentum observable. 

\subsection{Thermal Bath with temperature gradient}
\label{sec:tempgrad}
We now look at the scenario, where there is a temperature gradient.

Define $h(z)$ and $l(z)$ as follows:
\begin{equation}
	h(z) =  exp \left (-\frac{1}{2}\int \frac{f(z)}{4\sqrt{V(z)}} dz \right ).
\end{equation}
\begin{equation}
	\label{eq:li}
	l_i(z) = h(z)u_i(z), i=1,2.
\end{equation}
Equation~\ref{eq:li} indicates that the function $h(z)$ becomes then an exponential decay factor for $u_i(z)$, and depends only on the curvature dependent factor, $f(z)$.
The set of equations \ref{eq:secondorder}, can then be cast in the form,
\begin{eqnarray}
\label{eq:leq}
\nonumber	\frac{l_1^{''}}{l_1} = - {(-k_1k_2 + sk_2^{'} - m^2) - h^{''}(z))}\\
	\frac{l_2^{''}}{l_2} = - {(-k_1k_2 - sk_1^{'} - m^2) - h^{''}(z))}.
\end{eqnarray}
It can be observed that this is of a similar form as Eq.~\ref{eq:nograd}
The exact solution to the above has to be determined numerically, but if the temperature gradient is very small, one can further analyze Eq.~\ref{eq:leq} analytically.
\subsubsection{Small temperature gradient}
The temperature gradient being small implies that 
	the derivative, $\frac{\partial}{\partial z} \left \{-{(-k_1k_2 + sk_2^{'} - m^2) - h^{''}(z)} \right \} \sim 0$.
It can be seen that,
\begin{itemize}
	\item  $\frac{l_1^{''}}{l_1}$ = $\left \{-{(-k_1k_    2 + sk_2^{'} - m_g^2) - h^{''}(z)} \right \} = \left \{ (E^2 - m_g^2) - \frac{C^2}{V(z)} - s\frac{\partial}{\partial z}(\frac{C}{\sqrt{V(z)}}) + i\frac{2EC}{\sqrt{V(z)}} \right \}$ 
	\item $\frac{l_2^{''}}{l_2}$ =  $\left \{-{(-k_1k_    2 - sk_1^{'} - m_g^2) - h^{''}(z)} \right \} = \left \{ (E^2 - m_g^2) - \frac{C^2}{V(z)} + s\frac{\partial}{\partial z}(\frac{C}{\sqrt{V(z)}}) + i\frac{2EC}{\sqrt{V(z)}} \right \}$ 
	\item From the above two expressions for $\frac{l_1^{''}}{l_1}$ and $\frac{l_2^{''}}{l_2}$, it can be observed that the factor that is different for  $\frac{l_1^{''}}{l_1}$ and $\frac{l_2^{''}}{l_2}$, (and thus eventually  $\frac{u_1^{''}}{u_1}$ and $\frac{u_2^{''}}{u_2}$), is the term $s\frac{\partial}{\partial z}(\frac{C}{\sqrt{V(z)}})$. This term causes the two spinor functions $u_1$ and $u_2$ to have somewhat different derivatives, which implies that the Dirac spinor exists in a non-inertial frame. 
\end{itemize}
In the small gradient approximation, if  $\frac{\partial}{\partial z}(\frac{C}{\sqrt{V(z)}}) \sim 0$ 
	and $\frac{\partial}{\partial z} \left \{-{(-k_1k_2 + sk_2^{'} - m^2) - h^{''}(z)} \right \} \sim 0$,
one can define $q(z)$, to be the complex 3-momentum, where $q(z)^2 = -{(-k_1k_2 + sk_2^{'} - m^2) - h^{''}(z)} \approx \left \{ (E^2 - m_g^2) - \frac{C^2}{V(z)} +  \frac{i2EC}{\sqrt{V(z)}} \right \}$.
  The imaginary part of q(z), i.e. Imag(q(z)) gives the 3-momentum of the Dirac fermion, while the $re(q(z)) + h^{-1}(z)$ determines the total decay rate. $Imag(q(z))$ is given by
  \begin{equation}
  \label{eq:mom_obs}
	  Imag(q(z)) = \left ( \sqrt{\frac{1}{2} \left ( -Real(q^2) + |q^2|) \right )} \right )
  \end{equation}
	   This could then be interpreted as the "instantaneous" 3-momentum. Alternatively, $\frac{1}{Imag(q(z))}$ can be seen as the "instantaneous" de-Broglie wavelength inside the thermal bath. 
  The change in the 3-momentum leads to a time lag as the fermion passes through the thermal bath, which can then be measured.

  It can also be inferred that, if the temperature gradient and the initial momentum of the fermion are in perpendicular directions, then the change in the momentum along the direction of the temperature gradient would lead to curving of the fermion path.

  The curving of the fermion path is reminiscent of the geodesic equations in Sec.~\ref{sec:geodesic}, albeit in a quantum mechanical context.
\subsubsection{Large temperature gradient}
In this case, $e^{(\int q(z)dz)}$ would not solve Eq.\ref{eq:leq}. The solution has to be obtained numerically. 
Secondly, $\frac{l_2^{''}}{l_2} - \frac{l_1^{''}}{l_1} = 2s\frac{\partial }{\partial z} \frac{C}{\sqrt{V(z)}}$ and thus, $\frac{l_1^{''}}{l_1} \ne \frac{l_2^{''}}{l_2}$. For the above reasons, it is not straightforward to determine the 3-momentum observable. We now attempt to interpret the solution $l_1(z)$ and $l_2(z)$ obtained numerically. Define,
   		\begin{math}
			l(z) = 
		\left( \begin{array}{c}
			l_1\zeta_0\\
			l_2\zeta_0\\
		\end{array} \right ) 
   		\end{math}
			and,
   		\begin{math}
			u_p = 
		\left( \begin{array}{c}
			\sqrt(p.\sigma)\zeta_0\\
			\sqrt(p.\bar{\sigma})\zeta_0\\
		\end{array} \right ), 
   		\end{math}
		with $\sigma$ being the Pauli spin matrices, and $\zeta_0$ defined previously as part of Eq.~\ref{eq:spinor_u}.

		Fourier decompose $l(z)$ as $\int a_p u_p e^{ipz} dp$, where;
		\begin{equation}
			a_p = \frac{1}{m_g}\int u_p^t\gamma^0 l(z) e^{-ipz}dz
		\end{equation}
		The decomposition allows the wavefunction $l(z)$ to be interpreted  as a superposition of several fermionic waves of 3-momentum $p$, and amplitude $a_p$. 
	%	This seems akin to the Unruh radiation.
		However, a more general field theoretic approach would need to be followed in order to interpret $a_p$ as creation or annihilation operators. 
\subsection{Numerical analysis and Basic Setup}
\label{sec:numbase}
   A possible setup would be when a fermionic particle, say electron is fired through a plasma and is detected on the other side of the plasma. Both the source and detector are in vacuum, i.e. zero temperature, and the high temperature plasma is sandwiched in-between.
\begin{figure}
\includegraphics[width = 80mm,height = 80mm]{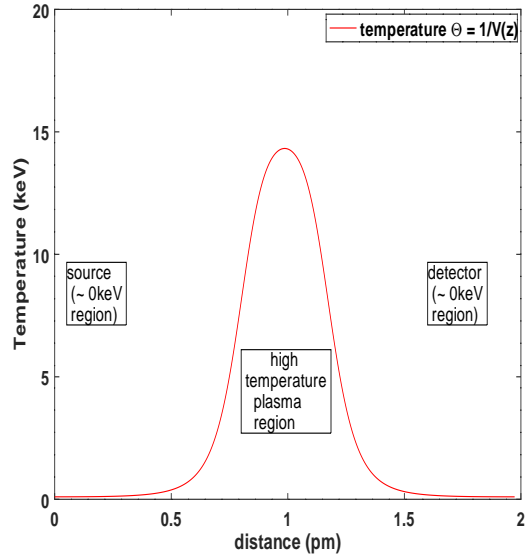}
	\caption{Plot of $\Theta(z)$ for Matsubara frequency $C = \frac{\pi}{\beta_T(0)}$.}
\label{fig:setup}
\end{figure}
   Figure~\ref{fig:setup} shows the experimental setup. The temperature is Gaussian-like (but not Gaussian). The temperature $\Theta$ and its derivative,  both approach 0 on either side, near the source and detector.
   The wave equation near the source and detector would then asymptotically tend to be the vacuum solution of the Dirac equation.
   The temperature curve $\Theta$, depicted in Fig.~\ref{fig:setup} is constructed using $V(z)$ as follows. Define, $\frac{\partial V(z)}{\partial z}$ as follows:
   \begin{eqnarray}
%	   \frac{dV_1}{dz} = -(\frac{scale}{\sqrt{2\pi\sigma_1^2}} exp(-\frac{(z - z_1).^2}{2\sigma_1^2};
%	   \frac{dV_2}{dz} = (\frac{scale}{\sqrt{2\pi\sigma_2^2}} exp(-\frac{(z - z_2).^2}{2\sigma_2^2};
%	   \frac{dV}{dz} = V_init(\frac{dV_1}{dz} + \frac{dV_2}{dz};
	   \nonumber	   \frac{dV}{dz} = V_{init} \Big (  -(\frac{1}{\sqrt{2\pi\sigma_1^2}} exp(-\frac{(z - z_1)^2}{2\sigma_1^2} \\
	   +  (\frac{1}{\sqrt{2\pi\sigma_2^2}} exp(-\frac{(z - z_2)^2}{2\sigma_2^2} \Big ),
   \end{eqnarray}
   One then gets $V(z)$ and $\Theta(z)$, given by
   \begin{eqnarray}
	   V(z) = \int_{z0}^{zl} \frac{dV}{dz} dz,\\
	   \Theta(z) = \frac{1}{V(z)}.
   \end{eqnarray}
	   The temperature, $\Theta(z)$, for $\sigma_1 = \sigma_2 = 0.81$, and ($z_1, z_2$)  = ($0.39460pm, 1.5784pm$), is depicted in Fig.~\ref{fig:setup}.
	   This construction ensures that both $V(z)$ and $\frac{dV}{dz}$ are nearly flat near the boundaries, allowing both $\Theta$ and curvature, $f(z)$, to be flat and small near the boundary.
	   The effective mass of the electron taken $=m =m_g + \gamma^0E$, where $m_g = 0.511MeV$ and $E = 5MeV$, for simulation.

 A linearly spaced grid of fifty thousand points was chosen for simulation and ranged from $z=z0=0$ to $z=zl = 10MeV^{-1} \approx 1.973pm$. 
	The initial conditions for simulations were estimated as:
	$u_1^{sim}(zl) \approx u_1^{flat}(zl)$, and
	$u_1^{sim'}(zl) \approx u_1^{flat'}(zl)$.

\begin{figure}
\includegraphics[width = 80mm,height = 80mm]{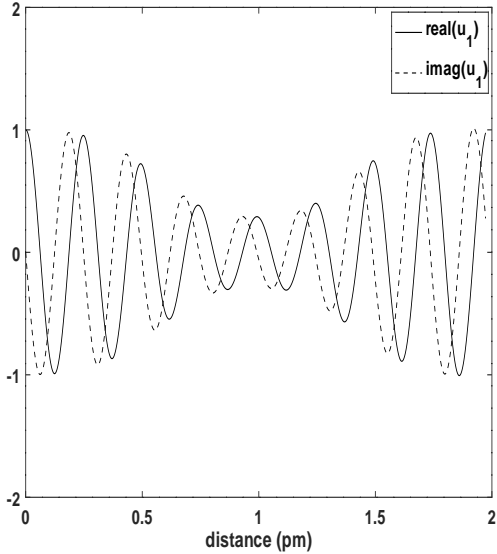}
	\caption{Plot of the wavefunction $l_1(z)$ for Matsubara frequency $C = \frac{\pi}{\beta_T(0)}$.}
\label{fig:l1}
\end{figure}

\begin{figure}
\includegraphics[width = 80mm,height = 80mm]{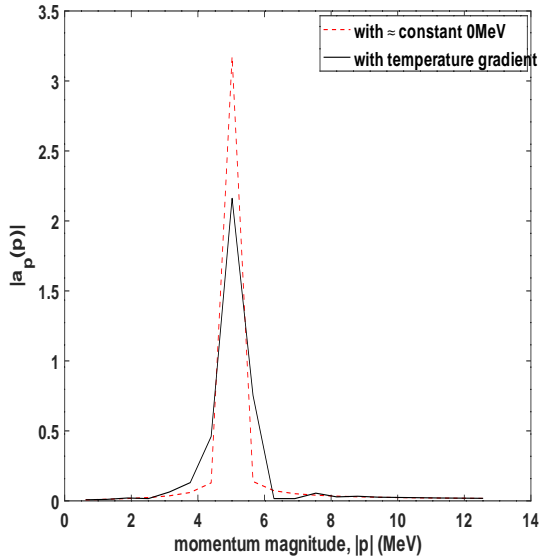}
	\caption{Fourier spectrum of the wavefunction $l_1(z)$ for no temperature gradient and large temperature gradient. Matsubara frequency $C = \frac{\pi}{\beta_T(0)}$.}
\label{fig:Fl1}
\end{figure}

	   We look at the case of the general solution which has been obtained numerically for a large temperature gradient. Figure, \ref{fig:l1} depicts the waveforms $Real(l_1)$ and $Imag(l_1)$ .
Figure ~\ref{fig:Fl1} depicts $|a_p(p)|$ for nil temperature gradient and large temperature gradient. The spreading of the spectrum can be interpreted as the superposition of additional momentum states. 

\begin{figure}
\includegraphics[width = 80mm,height = 80mm]{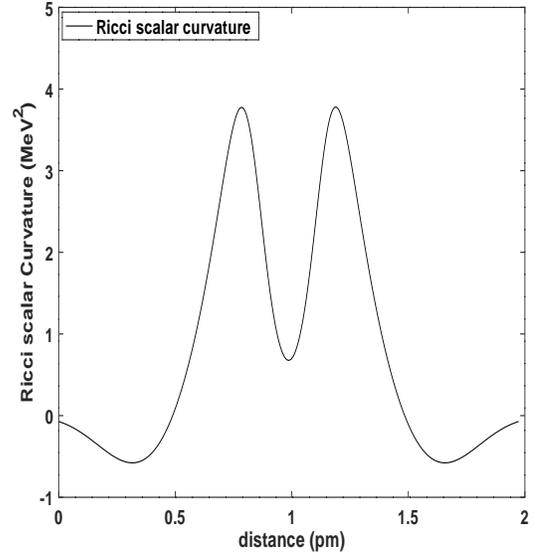}
	\caption{Plot of the Ricci scalar curvature at various values of $z$.}
\label{fig:ricci}
\end{figure}

	The Ricci scalar curvature, $R(z)$, corresponding to the temperature curve in Fig.~\ref{fig:setup} is plotted in Fig. \ref{fig:ricci}.

\section{Klein Gordon Particle}
\label{sec:kleingordon}
\subsection{Formulation}
The Klein Gordon equation for a massless particle would be
\begin{equation}
	\label{eq:kg}
%	(D_t^2 - D_x^2 - D_y^2 - D_z^2 + m^2 + U)A^{\nu} = 0,
	(D_t^2 - D_x^2 - D_y^2 - D_z^2 + U)A^{\nu} = 0,
\end{equation}
where, $D_{\mu}A^{\nu} = \partial_{\mu}A^{\nu} + \Gamma_{\mu\alpha}^{\nu}A^{\alpha}$, $t = i\beta$ and $\Gamma_{\mu\alpha}^{\nu}A^{\alpha}$ are the Christoffel coefficients.
$U$ is a phenomenological potential. As an example, in order to model the dispersion relation in an unmagnetized plasma, take $U = \omega_p^2 - k^2$, where $\omega_p$ is the plasma frequency~\cite{plasma3},
%$= \sqrt{\frac{n_e e^2}{\epsilon_0 m}}$~\cite{plasma3}, 
and $k$ is the incoming particle wave number.
%plasma3_reference Fitzpatrick:, Chapter 5, Sec. 5.7 (Waves in Unmagnetized Plasmas), page 131, Eg. 5.73
The dispersion relation indicates the frequency of  the electromagnetic (EM) wave, which would be able to propagate through the plasma, and those which would not be able to.
$U$, is reminiscent of the effective mass, $m = m_g + \gamma^0E$, of a fermion as mentioned in Sec.~\ref{sec:dirac}.
In case of a photon, the temperature becomes defined only in very high fields where the photon-photon interaction is non-zero.
Summarizing, $U$ determines the interaction between the incoming photons and the plasma ions, while the temperature is determined by the photon-photon interaction inside the plasma. 
Since photon-photon interaction is very small, the photon temperature may play only a very minor role as compared to the dispersion relation.
For a massless particle, propagating along the z-direction, $A^0=A^3=0$. The equation of motion then becomes,
\begin{equation}
	\frac{-1}{2V}A^{i'}V' - \frac{1}{V}\frac{\partial^2 A^i}{ \partial \beta^2} - A^{i''} = 0,  i = 1,2
\end{equation}
In flat space-time, i.e. when $V'=0$, the solution can be seen to be of the form
$A^i = e^{-iC\beta}e^q$, with $q = \pm\sqrt{\frac{C^2}{V(0)} + U}$, and $C = \frac{2n\pi}{\beta}$, with $n = 0,\pm 1,....$. 
For negligible temperature, i.e., $\beta \rightarrow \infty$, ($\equiv V(0) \rightarrow \infty$) this correctly reproduces the dispersion relation $q^2 = \omega_p^2 - k^2$. 
%Fitzpatrick:, Chapter 5, Sec. 5.7 (Waves in Unmagnetized Plasmas), page 131, Eg. 5.73
This dispersion relation indicates that if $k > \omega_p$, the photon would traverse the plasma. But, if $k < \omega_p$, then $q$ becomes real, and the photon would decay. 
In curved space-time, i.e. when $V' \ne 0$, we take the solution to be of the form,
\begin{equation}
	A^i = e^{-ig_{00}p^0\beta}A(z),
\end{equation}
where $A(z)$ is some function of $z$, and is determined numerically.
We do not expect Eq. \ref{eq:kg} to describe gluons, since, the three gluon vertex and four gluon vertex would alter the gluon propagator significantly, and a field theoretic description would be essential.

   One significant observation from Eq.~\ref{eq:kg}, would be when $\omega_p \approx k$, In this case, the potential $U$ becomes negligible, and the effects of the wavefunction would be purely due to the temperature resulting from photon-photon interaction. The geodesic Eq.~\ref{eq:xzcurve} implies that if there is a gradient in the photon density leading to a gradient in the photon temperature in at least a 2D plane, then the incoming photon can follow a curved path.
   In order to investigate the effect of curvature further, we assume that $\omega_p \approx k$, which effectively makes $U\approx 0$. Under this assumption we analyze the wavefunction in Sec.~\ref{sec:kgboundary}
\subsection{Boundary conditions and Observables}
\label{sec:kgboundary}
\begin{figure}
\includegraphics[width = 80mm,height = 80mm]{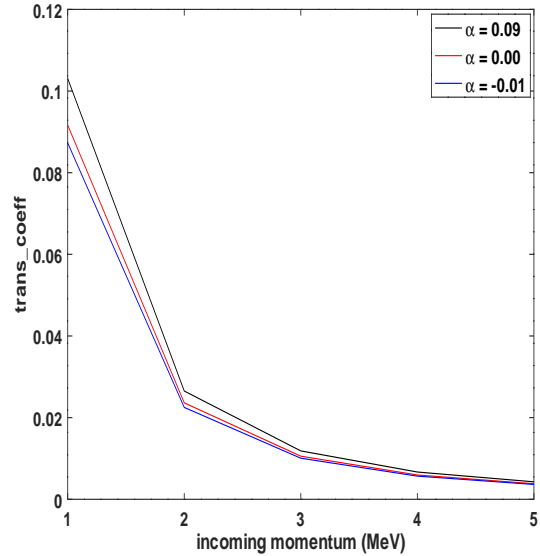}
	\caption{Transfer coefficients for various values of $\alpha$.} 
\label{fig:kg_transcoeff}
\end{figure}

Similar to the case of the Dirac spinor, 
we assume that the massless Klein Gordon particle (eg. photon), is propagating in vacuum, then enters into a plasma of width $a$, and then leaves again into a vacuum. This allows us to interpret the effect of curvature in terms of known observables in vacuum.
We take $V(z) = \frac{1}{(1+\alpha z)^2}$, between $z=0$ and $z=zl$ and $V(z)=1$ elsewhere. Also assume the plasma frequency to be constant throughout (i.e. independent of $z$). This could be a valid approximation if $\alpha$ is small.
One needs to solve the boundary conditions around $z=0 and z=zl$.
Let the spatial component of the incoming wave be $P_1e^{ikz} + P_2e^{-ikz}$, and the spatial component of the outgoing wavefunction be $Re^{ilz}$. The  spatial component of the wavefunction in the plasma is given by $B_1A_+(z) + B_2A_{-}(z)$. 
In the absence of temperature gradient the two solutions $A_+(z)$ and $A_{-}(z)$, would correspond to $q = \pm\sqrt{\frac{C^2}{V(0)}}$.
Equating the spatial component of the wavefunction and its derivative at the boundaries, helps determine the ratio $\frac{R^2}{P_1^2}$.
Inside the plasma, the equation of motion does not contain any time component in the imaginary time formalism. Thus, the time component is taken to remain unaffected. For a photon, this implies that the incoming and outgoing frequencies are the same. 
The effect of temperature gradient is depicted in Fig.~\ref{fig:kg_transcoeff}. 
The curvature of the Euclideanized spacetime leads to modification of the probability of photons transmitted.

\section{Conclusion} 
\label{sec:conclusion}
Starting with the help of a Polyakov loop correlator, we have propounded a mathematical equivalence between temperature gradient and curvature of Euclideanized spacetime. The variation in temperature is recast as a variation in the metric.
    The Dirac equation for a massive fermion is then solved in the resulting Euclideanized curved spacetime. The solution indicates an exponentially decaying wavefunction in the thermal bath. The spacetime curvature leads to modification in the decay rate.

 	It may be possible to stretch the mathematical equivalence to  regions where a characteristic length in the system becomes comparable to that of the curvature of  the Euclideanized spacetime. 
We have seen that, when the temperature gradient is small, i.e. $\frac{\partial}{\partial z}\left ( \frac{C}{\sqrt{V(z)}} \right )$ is small, the quantity $q_1^2 = \left | \frac{u_1^{''}}{u_1} \right |$, has the interpretation of square momentum (including the decay rate). 
Similarly, $q_2^2 = \left | \frac{u_2^{''}}{u_2} \right |$, provides the square momentum for the spinor $u_2$.
In Quantum systems in curved spacetime, an interesting region would be when the de Broglie wavelength $> \frac{1}{\sqrt{R}}$, where $R$ = Ricci scalar curvature. 
In a similar analogy, a region of interest could be when $(\frac{1}{q_1})^2 \ge \frac{1}{R}$ and $(\frac{1}{q_2}) \ge \frac{1}{R}$, or alternatively, when $q_1^2 \le R$ and  $q_2^2 \le R$.
This can happen when 
\begin{itemize}
\item The effective mass $m$  is very small. This can be seen to be effectively the case, when $E$ is small. 
\item The Matsubara frequency $C$ is large, or alternatively, one has a large $\beta_T$, and one is looking at large Matsubara frequencies. These conditions then boil down to having high mean temperature and relatively large temperature gradient. Large temperature gradients will lead to large $R$. 
\end{itemize}
	However large $C$ or large temperature gradient makes $q_i$ (i=1,2) less well defined. Thus it may be better to keep $E$ as small as possible so that $|q_i| \gg \sqrt{R}$.
In summary, regions where $|q_i| \gg \sqrt{R}$ (i= 1,2) and $|q_i| \le \sqrt{R}$, would be two regions where one might want to explore the Dirac equation in curved space. 
However, local thermal equilibrium may break down for large $R$,  rendering the region $|q_i| \le \sqrt{R}$, difficult to explore.

	%It is seen that the length scales shown in the Fig. \ref{fig:zeta1}, are in picometer range, which would still be formidable to reproduce in the lab. But at least the scales are not in the Planck scale. 
	%Future research may be able to come up with scenarios, where the distances involved are in nanometer range instead of picometer range. 
%	For example, the potential between muon and anti-muon in a QED plasma with a temperature gradient, may involve distances in nanometer range. 

\section{Acknowledgments} 
I would like to thank Prof. Michael E. Peskin (SLAC National Accelerator Lab.) for useful discussions.

\end{document}